\documentclass[seceq]{ptptex}
\usepackage{graphics}

\pubinfo{Vol. 127, No. 4, April 2012}
\title{Analysis of Popper's Experiment and its Realization}

\author{Tabish \textsc{Qureshi}\footnote{E-mail: tabish@ctp-jamia.res.in}}
\inst{Centre for Theoretical Physics\\ Jamia Millia Islamia,
New Delhi-110025, India.}

\abst{An experiment proposed by Karl Popper to test the standard interpretation
of quantum mechanics was realized by Kim and Shih.  We use a quantum
mechanical calculation to analyze Popper's proposal, and find a surprising
result for the location of the virtual slit.  We also analyze Kim
and Shih's experiment, and demonstrate that although it ingeniously
overcomes the problem of temporal spreading of the wave-packet, it is
inconclusive about Popper's test. We point out that another experiment
which (unknowingly) implements Popper's test in a conclusive way, has
actually been carried out.  Its results are in contradiction with
Popper's prediction, and agree with our analysis.}

\begin{document}
\maketitle

\section{Introduction}
The evidently nonlocal character of quantum mechanics has been a
source of discomfort right from the time of its inception. Einstein
Podolsky and Rosen, in their seminal paper, introduced a thought experiment,
which became famous as the EPR experiment, articulating the disagreement
of quantum theory with the classical notion of locality
\cite{epr}.

A lesser known experiment was proposed by Karl Popper, who
called it a variant of the EPR experiment, to test the
standard interpretation of quantum theory \cite{popper,popper1}.
Popper's proposed experiment consists of a source $S$ that can generate pairs
of particles traveling to the left and to the right along the $x$-axis. The
momentum along the $y$-direction of the two particles is entangled in such a
way so as to conserve the initial momentum at the source,
which is zero. There are two slits, one each in the paths of the two particles.
Behind the slits are semicircular arrays of detectors which can detect the
particles after they pass through the slits (see Fig.~1).

\begin{figure}
\centerline{\resizebox{4.5in}{!}{\includegraphics{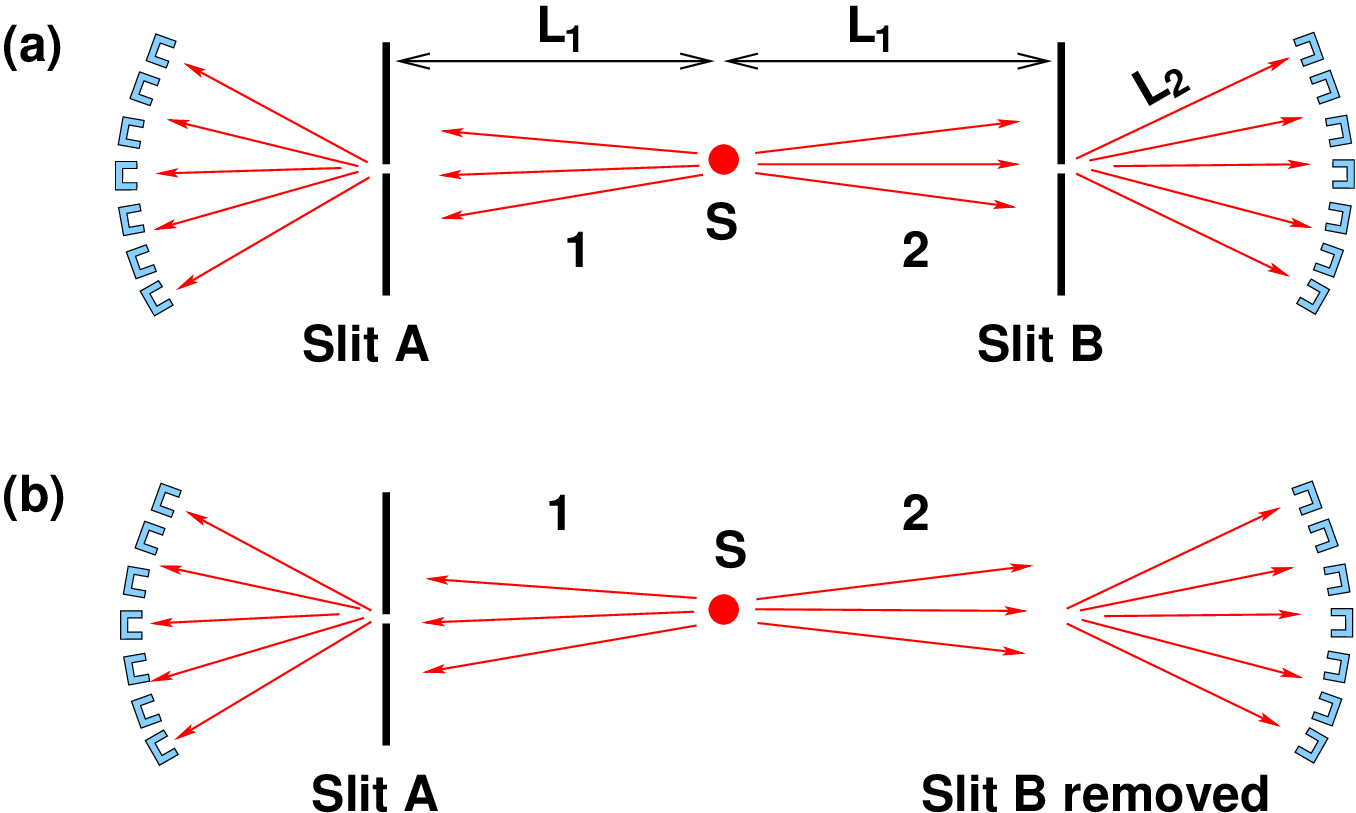}}}
\caption{Schematic diagram of Popper's thought experiment. (a) With both 
slits, the particles are expected to show scatter in momentum. (b) By removing
slit B, Popper believed that the standard interpretation of
quantum mechanics could be tested.}
\end{figure}

Being entangled in momentum space implies that in the absence of the two
slits, if a particle on the left is measured to have a momentum $p$, the particle
on the right will necessarily be found to have a momentum $-p$. One can
imagine a state similar to the EPR state, 
$\psi(y_1,y_2) = \!\int_{-\infty}^{\infty}e^{ipy_1/\hbar} e^{-ipy_2/\hbar}dp$.
As we can see, this state also implies that if a particle on the left is
detected at a distance $y$ from the horizontal line, the particle on the right
will necessarily be found at the same distance $y$ from the horizontal line.
It appears, however, that a hidden assumption in Popper's setup is that
the initial spread in momentum of the two particles is not very large.
Popper argued that because the slits localize the particles to a narrow
region along the $y$-axis,
they experience large uncertainties in the $y$-components of their momenta.
This larger spread in the momentum will show up as particles being
detected even at positions that lie outside the regions where particles
would normally reach based on their initial momentum spread. This is generally
understood as a diffraction spread.
 
Popper suggested that slit B be made very large (in effect, removed).
In this situation, Popper argued
that when particle 1 passes through slit A, it is localized to
within the width of the slit. 
He further argued that the standard interpretation of quantum mechanics
tells us that if particle 1 is localized in a small region of space, particle
2 should become similarly localized, because of entanglement.
The standard interpretation says that if one has knowledge about the
position of particle 2, that should be sufficient to cause a spread in
the momentum, just from the Heisenberg uncertainty principle. Popper said
that he was inclined to believe that there will be no spread in the particles
at slit B, just by putting a narrow slit at A.

Popper's proposed experiment came under lot of attention, especially because
it represented an argument which was falsifiable, an experiment which could
actually be carried out \cite{sudbery,sudbery2,krips,collet,storey,redhead,
nha,peres,hunter,sancho,tqijqi}. The experiment was realized in 1999 by
Kim and Shih using a
spontaneous parametric down conversion (SPDC) photon source to generate
entangled photons\cite{shih}.  They did not
observe an extra spread in the momentum of particle 2 due to particle
1 passing through a narrow slit. In fact, the observed momentum spread
was narrower than that contained in the original beam. This observation
seemed to imply that Popper was right. Short criticized Kim and
Shih's experiment, arguing that because of the finite size of the source,
the localization of particle 2 is imperfect,\cite{short} which leads to
a smaller momentum spread than expected.
It has been shown earlier that according to standard interpretation of
quantum mechanics, particle 2 cannot have any extra momentum spread
\cite{tqajp}. An extra momentum spread in particle 2 would also imply
a possibility of sending a faster-than-light signal, which is known
to be impossible \cite{gerjuoy}. However, a good explanation of the
results of Kim and Shih's experiment is still lacking.
In this paper, we do a rigorous analysis of the dynamics of an entangled
state, passing through a slit. We will show that this is necessary to
meaningfully interpret the results of Kim and Shih's experiment.

\section{Dynamics of entangled particles}

\subsection{The Entangled State}
First thing one must recognize is that in a real SPDC source, the correlation
between the signal and idler photons is not perfect. Several factors like
the finite width of the nonlinear crystal, finite waist of the pump beam and
the spectral width of the pump, play important role in determining how good
is the correlation \cite{spdc}.
Therefore, we assume the entangled particles, when they start out at the
source, to be in a state which has the following form,
\begin{equation}
\psi(y_1,y_2) = C\!\int_{-\infty}^\infty dp
e^{-p^2/4\sigma^2}e^{-ipy_2/\hbar} e^{i py_1/\hbar}
e^{-{(y_1+y_2)^2\over 4\Omega^2}}, \label{state}
\end{equation}
where $C$ is a normalization constant. The
$e^{-(y_1+y_2)^2/4\Omega^2}$ term, apart from making the state (\ref{state})
normalized, also restricts the spread in both $y_1$ and $y_2$.
The state (\ref{state}) is fairly general, except that we use Gaussian
functions.

In order to study the evolution of the particles as they travel towards
the slits, we will use the following strategy. Since the motion along
the x-axis is unaffected by the entanglement of the form given by 
(\ref{state}), we will ignore the x-dependence of the state. We will
assume the particles to be traveling with an average momentum $p_0$,
so that after a known time, particle 1 will reach slit A.
So, motion along the $x$-axis is ignored, but is implicitly included in the
time evolution of the state. Integration over $p$ can be carried out in
(\ref{state}), to yield the normalized state of the particles at time $t=0$,
\begin{equation}
\psi(y_1,y_2,0) = \sqrt{\sigma\over\pi\hbar\Omega} e^{-(y_1-y_2)^2\sigma^2/\hbar^2}
e^{-(y_1+y_2)^2/4\Omega^2} .  \label{newstate}
\end{equation}
The uncertainty in the momenta of the two particles given by
$\Delta p_{1y}= \Delta p_{2y}=\sqrt{\sigma^2 + {\hbar^2/4\Omega^2}}$.
The position uncertainty of the two particles is
$\Delta y_1 = \Delta y_2 = {1\over 2}\sqrt{\Omega^2+\hbar^2/4\sigma^2}$.
While the constants $\Omega$ and $\sigma$ can take arbitrarily values,
the form of (\ref{newstate}) makes sure that the uncertainty relation
is always respected.
Let us assume that the particles travel for a time $t_1$ before particle 1
reaches slit A. The state of the particles after a time $t_1$ is given by
\begin{equation}
\psi(y_1,y_2,t_1) = \exp\left(-{i\over\hbar}\mathbf{H}t_1\right)\psi(y_1,y_2,0) 
\end{equation}
The Hamiltonian $\mathbf{H}$ being the free particle Hamiltonian for the two
particles, the state (\ref{newstate}), after a time $t_1$ looks like
\begin{equation}
\psi(y_1,y_2,t_1) = {1\over \sqrt{\pi(\Omega+{i\hbar t_1\over m\Omega})
({\hbar\over \sigma} + {4i\hbar t_1\over m\hbar/\sigma})}}
\exp\left({-(y_1-y_2)^2\over
{\hbar^2\over\sigma^2} + {4i\hbar t_1\over m}}\right)
 \exp\left({-(y_1+y_2)^2\over 4(\Omega^2+ {i\hbar t_1\over m})}\right).
\label{statet1}
\end{equation}

\subsection{Effect of slit A}
At time $t_1$ particle one passes through the slit. We may assume that the
effect of the slit is to localize the particle into a state with position
spread equal to the width of the slit.
Let us suppose that the wave-function of particle 1 is reduced to
\begin{equation}
\phi_1(y_1) = \frac{1}{(\epsilon^2\pi/2)^{1/4} } e^{-y_1^2/\epsilon^2}.
\label{phi1}
\end{equation}
In this state, the uncertainty in $y_1$ is given by
	$\Delta y_1 = \epsilon/2$.
The measurement destroys the entanglement, but
the wave-function of particle 2 is now known to be:
\begin{eqnarray}
\phi_2(y_2) &=& \!\int_{-\infty}^\infty \phi_1^*(y_1) \psi(y_1,y_2,t_1) 
dy_1 \label{phi2f}
\end{eqnarray}
We had argued earlier \cite{tqijqi} that mere presence of slit A does not
lead to a reduction of the state of the particle. While strictly speaking,
this is true, one would notice that if one assumes that the wave-function
is not reduced, part of the wave function of particle 1 passes through the
slit, and a part doesn't pass. The part which passes through the slit, is
just $\phi_1(y_1)\phi_2(y_2)$. By the linearity of Schr\"odinger equation,
each part will subsequently evolve independently, without affecting the other.
If we are only interested in those pairs where particle 1 passes through
slit A, both the views lead to identical results. Thus, whether one believes
that the presence of slit A causes a collapse of the wave-function or not,
one is led to the same result.
 
The state of particle 2, given by (\ref{phi2f}), after normalization, has
the explicit form
\begin{eqnarray}
\phi_2(y_2) &=& \left({\Gamma+\Gamma^*\over \pi\Gamma^*\Gamma}\right)^{1/4}
\exp\left(-{y_2^2\over\Gamma}\right), \label{phi2}
\end{eqnarray}
where 
\begin{equation}
\Gamma = \frac{\epsilon^2+2i\hbar t_1/m+{\hbar^2/\sigma^2\over
1 + \hbar^2/(4\sigma^2\Omega^2) }}{1+{\epsilon^2+2i\hbar t_1/m\over\Omega^2 +
\hbar^2/4\sigma^2}} + {2i\hbar t_1\over m} .\label{dy2}
\end{equation}
The above expression simplifies in the limit $\Omega \gg \epsilon$,
$\Omega \gg \hbar/2\sigma$. In this limit, (\ref{phi2}) is a Gaussian
function, with a width $\sqrt{\epsilon^2+\hbar^2/\sigma^2 +
{16\hbar^2 t_1^2/m^2 \over\epsilon^2+\hbar^2/\sigma^2}}$. In the limit
$\hbar/\sigma \to 0$, the correlation between the two particles is expected
to be perfect. One can see that even in this limit, localization of
particle 2 is not perfect. It is localized to a region of width
$\sqrt{\epsilon^2 + {16\hbar^2 t_1^2/m^2 \over\epsilon^2}}$. So, Popper's
thinking that an initial EPR like state implies that localizing particle
1 in a narrow region of space, after it reaches the slit, will lead to a
localization of particle 2 in a region as narrow, is not correct.

Once particle 2
is localized to a narrow region in space, its subsequent evolution should
show the momentum spread dictated by (\ref{phi2}).
The uncertainty in the momentum of particle 2 is now given by
\begin{eqnarray}
\Delta p_{2y} &=& \int_{-\infty}^{\infty}\phi_2^*(y_2)(\mathbf{p}-\langle\mathbf{p}
\rangle)^2\phi_2(y_2)dy_2\nonumber\\
&=& {\sqrt{2}\hbar\over\sqrt{\Gamma + \Gamma^*}} \nonumber\\
&\approx& {\sigma\over \sqrt{1 + \left({\sigma\epsilon\over\hbar}\right)^2
+ \left({2\sigma t_1\over m\Omega}\right)^2}} ,
\label{dp2n}
\end{eqnarray}
where the approximate form in the last step emerges for the realistic
scenario $\Omega \gg \epsilon$, $\Omega \gg \hbar/2\sigma$ and
$\Omega^2\gg 2\hbar t_1/m$.
Clearly, the momentum spread of particle 2 is always less than that present in
the initial state, which was $\sqrt{\sigma^2 + {\hbar^2/4\Omega^2}}
\approx \sigma$.

\subsection{The Virtual Slit}
After particle 1 has reached slit A, particle 2 travels for a time $t_2$
to reach the array of detectors. The state of particle 2, when it reaches
the detectors, is given by
\begin{equation}
\phi_2(y_2,t_2) = \left({\Gamma+\Gamma^*\over \pi\Gamma'^*\Gamma'}\right)^{1/4}
\exp\left(-{y_2^2\over\Gamma'}\right), \label{phi2d}
\end{equation}
where $\Gamma' = \Gamma + 2i\hbar t_2/m$. In the limit $\Omega \gg \epsilon$,
$\Omega \gg \hbar/2\sigma$, (\ref{phi2d}) assumes the form
\begin{equation}
\phi_2(y_2,t_2) \approx \left({2\over\pi}\right)^{1/4}
\left(\sqrt{\epsilon^2+{\hbar^2\over\sigma^2}}
+{2i\hbar(2t_1+t_2)\over m\sqrt{\epsilon^2+{\hbar^2\over\sigma^2}}}\right)^{-1/2}
\exp\left(-{y_2^2\over \epsilon^2+{\hbar^2\over\sigma^2}
+{2i\hbar(2t_1+t_2)\over m}}\right), \label{phi2final}
\end{equation}
Equation (\ref{phi2final}) represents a Gaussian state, which has undergone
a time evolution. But this form implies that particle 2 started out as
Gaussian state, with a width $\sqrt{\epsilon^2+\hbar^2/\sigma^2}$, and traveled
for a time $2t_1+t_2$. But the time $2t_1+t_2$ corresponds to the particle
having traveled a distance $2L_1+L_2$, which is the distance between slit A
and the detectors behind slit B. This is very strange because particle 2
never visits the region between the source and slit A. If particle 1 were
localized right at the source, the width of the localization of particle 2
would have been
$\sqrt{\epsilon^2+\hbar^2/\sigma^2}$ (for large $\Omega$). {\em So, the virtual
slit for particle 2 appears to be located at slit A, and not at slit B.}
However, the width of the virtual slit will be more than the real slit A,
and the diffraction observed for particles 1 and 2 will be different.

\section{Kim and Shih's experiment}

\subsection{Width of the observed pattern}

In order to use the results obtained in the preceding section, we will recast
them in terms of the d`Broglie wavelength of the particles. In this
representation, (\ref{phi2final}) has the form
\begin{equation}
\phi_2(y_2,t_2) \approx \left({2\over\pi}\right)^{1/4}
\left(\sqrt{\epsilon^2+{\hbar^2\over\sigma^2}}
+{i\lambda(2L_1+L_2)\over\pi\sqrt{\epsilon^2+{\hbar^2\over\sigma^2}}}\right)^{-1/2}
\exp\left({-y_2^2\over \epsilon^2+{\hbar^2\over\sigma^2}
+{i\lambda(2L_1+L_2)\over\pi}}\right), 
\end{equation}
where $\lambda$ is the d`Broglie wavelength associated with the particles.
For photons, $\lambda$ will represent the wavelength of the photon. For
convenience, we will use a rescaled wavelength $\Lambda=\lambda/\pi$. The
probability density distribution of particle 2 at the detectors behind slit B,
is given by $|\phi_2(y_2,t_2)|^2$, which is a Gaussian with a width equal to
\begin{equation}
W_2 = \sqrt{\epsilon^2+{\hbar^2\over\sigma^2}
+{4\Lambda^2(2L_1+L_2)^2\over\epsilon^2+\hbar^2/\sigma^2}}. \label{width}
\end{equation}

\begin{figure}[h]
\centerline{\resizebox{4.5in}{!}{\includegraphics{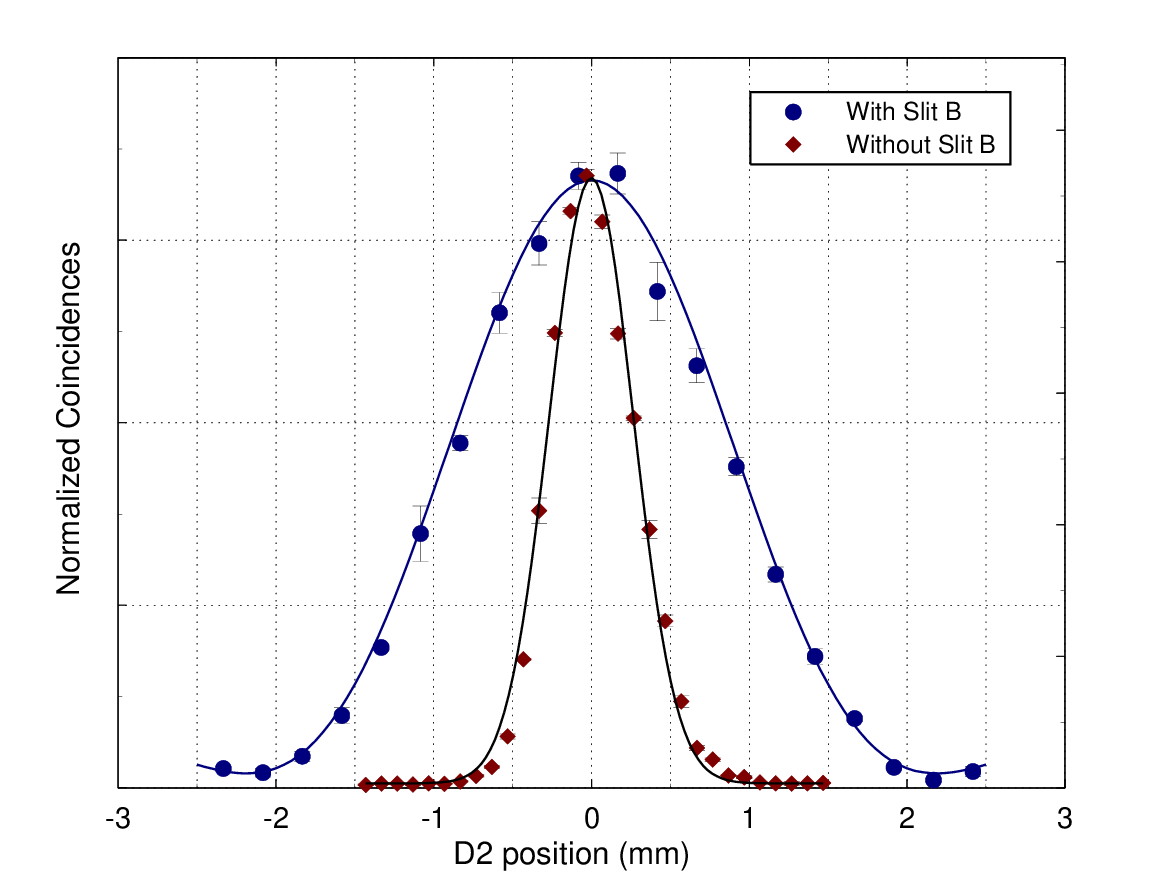}}}
\caption{Results of the photon experiment by Kim and Shih, \cite{shih} aimed
at realizing Popper's proposal. The diffraction pattern in the absence of
slit B (diamond symbols) is much narrower than that in the presence of a
real slit (round symbols).}
\end{figure}

Let us now look at the experimental results of Kim and Shih. Equation
(\ref{width}) should represent the width of the observed pattern in their
experiment (see Fig. 2).  They observed
that when the width of slit B is 0.16 mm, the width of the diffraction pattern
(at half maximum) is 2 mm. When the width of slit A is  0.16 mm, but slit B
is left wide open, the width of the diffraction pattern
is 0.657 mm.
In a Gaussian function, the full width at half maximum
is related to the Gaussian width $W$ by
\begin{equation}
	W_{fwhm} = \log(2)W
\end{equation}
Using $W_2 = 0.657/\log(2)$ mm, $\lambda = 702$ nm and $2L_1 + L_2 = 2$ m,
we find $\sqrt{\epsilon^2+\hbar^2/\sigma^2} = 0.632$ mm. Assuming that
a rectangular slit of width 0.16 mm corresponds to a Gaussian width
$\epsilon = 0.11$ mm, the number  0.632 for $\sqrt{\epsilon^2+\hbar^2/\sigma^2}$
is unusually large. If the number 0.632 really pertains to 
$\sqrt{\epsilon^2+\hbar^2/\sigma^2}$, it would means that the effect of
imperfect correlation (represented by $\hbar/\sigma$) is much much larger
than the localization effect of the slit.  Clearly, something is amiss here.

A careful look reveals that the analysis we presented in the last section
applies to freely evolving entangled particle, while Kim and Shih's
setup also involves a converging lens. Thus, the photons are not
really free particles - their dynamics is affected by the lens. So,
our next task is to incorporate the effect of the lens in our calculation.

\subsection{Effect of converging lens}

We assume the effect of a converging lens of focal length $f$ to be the
following. If a Gaussian wave-packet of width $\sigma$ starts from a distance
$2f$ from the lens, it will spread due to time evolution as it reaches the
lens. The effect of lens is to have a unitary transformation on the
wave-packet such that in its subsequent dynamics, it narrows instead of
spreading, and comes back to its original width after a distance $2f$ from
the lens. Also, the observed width of the wavepacket, immediately after
emerging from the lens should be the same as that just before entering
the lens. In general, we can quantify the effect of the lens by a unitary
transformation of the form
\begin{eqnarray}
  \mathbf{U}_f {(\pi/2)^{-1/4}\over\sqrt{\sigma+{i\Lambda L\over\sigma}}} 
\exp\left({-y_1^2 \over \sigma^2+i\Lambda L}\right) &=& 
{(\pi/2)^{-1/4}\over\sqrt{\tilde{\sigma}+{i\Lambda (L-4f)\over
\tilde{\sigma}}}}\nonumber\\
&&\exp\left({-y_1^2 \over \tilde{\sigma}^2+i\Lambda (L-4f)}\right),
\label{lens} \nonumber\\
\end{eqnarray}
where $L$ is the distance the wave-packet, of an initial width $\sigma$, 
traveled before passing through the lens, and $\tilde{\sigma}$ is such that
it satisfies
\begin{equation}
\tilde{\sigma}^2+{\Lambda^2(L-4f)^2\over\tilde{\sigma}^2} =
\sigma^2+{\Lambda^2 L^2\over\sigma^2}.
\end{equation}
One can verify that if $L=2f$, the state emerging from the lens, given
by (\ref{lens}), after traveling a further distance $2f$, assumes the form
${(\pi/2)^{-1/4}\over\sqrt{\sigma}} \exp\left({-y_1^2 \over \sigma^2}\right)$.

\begin{figure}[h]
\centerline{\resizebox{5.0in}{!}{\includegraphics{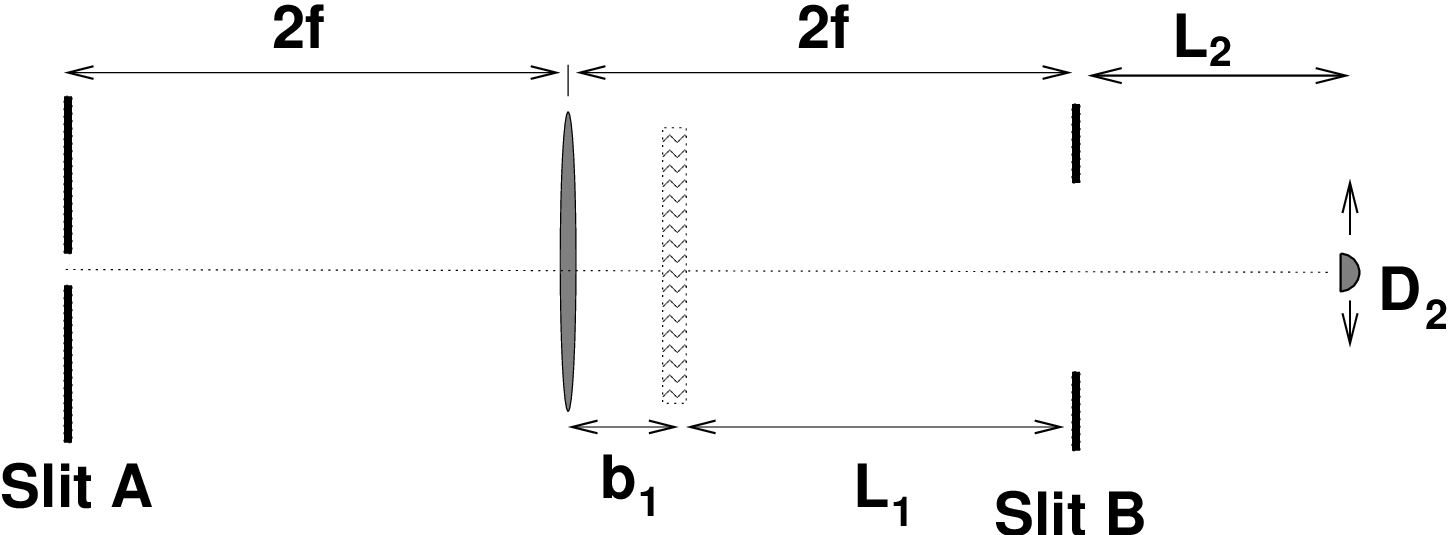}}}
\caption{Setup of the photon experiment by Kim and Shih, \cite{shih} aimed
at realizing Popper's proposal. Slit A is narrow while slit B is left wide
open.}
\end{figure}

In this scenario, we split the time $t_1$, taken by particle 1 to reach
slit A, into two parts: the time $t_{b1}$ taken to travel the distance $b_1$,
from the source to the lens, and the time $t_{2f}$ taken to travel the
distance $2f$, from the lens to slit A. So, the state of particle 2,
after a time $t_1$, conditioned on particle 1 having passed through slit A,
is given by
\begin{eqnarray}
\phi_2(y_2,t_1) &=& \int_{-\infty}^\infty \phi_1^*(y_1) \psi(y_1,y_2,t_1)
dy_1  \nonumber\\
&=& \int_{-\infty}^\infty \phi_1^*(y_1) e^{-{i\over\hbar}H_1t_{2f}}
\mathbf{U}_fe^{-{i\over\hbar}H_1t_{b2}} 
 e^{-{i\over\hbar}H_2t_1} \psi(y_1,y_2,0) dy_1
\end{eqnarray}
Similarly, one can write the state of particle 2 at a general time $t$,
conditioned on particle 1 having passed through slit A, as
\begin{eqnarray}
\phi_2(y_2,t) = \int_{-\infty}^\infty \phi_1^*(y_1)
e^{-{i\over\hbar}H_1t_{2f}} \mathbf{U}_fe^{-{i\over\hbar}H_1t_{b2}}
e^{-{i\over\hbar}H_2t}\psi(y_1,y_2,0) dy_1 \label{phi2t}
\end{eqnarray}
A word of caution is needed while interpreting (\ref{phi2t}). For a time
$t < t_1$, the two particle state is actually an entangled state, which
renders any attempt to write the wave-function of just particle 2,
meaningless. For such a purpose, normally one has to resort to mixed states.
However, if one were to calculate any quantity, including the probability
of finding particle 2 in a certain region of space, {\em conditioned on
particle 1 having passed through slit 1}, (\ref{phi2t}) will give the correct
result, even for a time before $t_1$.

For $\phi_1(y_1)$ given by (\ref{phi1}), the wave-function of particle 2,
at a time $t$, has the explicit form
\begin{eqnarray}
\phi_2(y_2) = \exp\left({- y_2^2 \over \epsilon^2 + {\hbar^2\over
\sigma^2}+i2\Lambda(b_1-2f)+2i\Lambda L}\right),
\end{eqnarray}
where $L$ is the distance traveled by the particle in time $t$ and $C$
is a constant necessary for normalization. For $L=2f-b_1$, $\phi_2(y_2)$
is a Gaussian with a width equal to $\sqrt{\epsilon^2 +
\hbar^2/\sigma^2}$, which is exactly the position spread of particle 2,
when it started out at the source. $L=2f-b_1$ corresponds to particle 2
being at slit B. Indeed, we see that because of the clever arrangement
of the setup in Kim and Shih's experiment, particle 2 is localized at
slit B to a region as narrow as its initial spread. Thus, the spreading
of the wave-packet because of temporal evolution, which would have been
present in Popper's original setup, has been avoided. So, in Kim and Shih's
realization, the virtual slit is indeed at the location of slit B. However,
its width is  larger than the width of the real slit.

Now one can calculate the width of the distribution of
particle 2, as seen by detector D2. In reaching detector D2, particle 2
travels a distance $L=L_1+L_2 = 2f-b_1 + L_2$. The width (at half maximum)
of pattern at D2 is now given by
\begin{equation}
W_2 = \sqrt{\epsilon^2+{\hbar^2\over\sigma^2}
+{4\Lambda^2L_2^2\over\epsilon^2+\hbar^2/\sigma^2}}. \label{widthnew}
\end{equation}
Contrasting this expression with (\ref{width}), one can explicitly see
the effect of introducing the lens in the experiment - basically, the length
$L_2$ occurs here in place of $2L_1+L_2$.
Using $W_2 = 0.657/\log(2)$ mm, $\lambda = 702$ nm and $2L_1 + L_2 = 2$ m,
we now find $\sqrt{\epsilon^2+\hbar^2/\sigma^2} = 0.236$ mm. Assuming that
a rectangular slit of width 0.16 mm corresponds to a Gaussian width
$\epsilon = 0.08$ mm (which gives the correct diffraction pattern width for
a {\em real} slit), we find $\hbar^2/\sigma^2 = 0.049~mm^2$. For a perfect
EPR state, $\hbar^2/\sigma^2$ should be zero. So, we see that for a real
entangled source, where correlations are not perfect, a small value of
$\hbar^2/\sigma^2 = 0.049~mm^2$, satisfactorily explains why the diffraction
pattern width is 0.657 mm, as opposed to the width of 2 mm for a real slit
of the same width. 

From the preceding analysis, it is clear that if $\hbar/\sigma$ were
zero, the diffraction pattern would be as wide as that for a real slit.
However, the smaller the quantity $\hbar/\sigma$, the more divergent is
the beam. This can be seen from (\ref{statet1}), which implies that
an initial width of the beam $\Delta y_2=\sqrt{\Omega^2+\hbar^2/4\sigma^2}$,
corresponds to a width $\sqrt{\Omega^2+{\Lambda^2 L^2\over\Omega^2}
+\hbar^2/4\sigma^2+{\Lambda^2 L^2\over\hbar^2/\sigma^2}}$, after particle
2 has traveled a distance $L$. Consequently, the width of the diffraction
pattern is never larger than the width of the beam, in the case of
diffraction from a virtual slit. Width of the beam here refers to the width
of the pattern obtained from {\em all} the counts, without any coincident
counting. Thus, no additional momentum spread can
ever be seen in Popper's experiment.
The conclusion is that although Kim and Shih correctly implemented Popper's
experiment through the innovative use of the converging lens, it is not
decisive about Popper's test of the Copenhagen interpretation, because of
imperfect correlation between the two photons.

\section{The Real Popper's Test}

The discussion in the preceding section implies that making the the
correlation
of the two entangled particles better, doesn't throw any new light on the
issue. However there is a way in which Popper's test can be implemented.
Popper states:\cite{popper}
\begin{quote}
``if the Copenhagen interpretation is
correct, then any increase in the precision in the measurement of our {\em
mere knowledge} of the particles going through slit B should increase their
scatter.''
\end{quote}
This view just says that if the
(indirect) localization of particle 2 is made more precise, the momentum
spread should show an increase. This could have easily been done in Kim
and Shih's experiment by gradually narrowing slit A, and observing the 
corresponding diffraction pattern. 

An experiment which (unknowingly) implements this idea, has actually been
performed,
although its connection to Popper's proposal has not been recognized. This is
the so-called ghost interference experiment by Strekalov et al \cite{ghost}.
In the single slit ghost interference experiment,
a SPDC source generates entangled photons and a single slit is put in the
path of one of these. There is a lone detector D1 sitting behind the single
slit, and a detector D2, in the path of the second photon, is scanned along
the y direction, after a certain distance. The only way in which this experiment
is different from the Popper's proposed experiment is that D1 is kept fixed,
instead of being scanned along y-axis or placed in front of a collection lens
as in \cite{shih}. Now, the reason for doing coincident
counting in Popper's experiment was to make sure that only those particles
behind slit B where counted, whose entangled partner passed through
slit A. This was supposed to see the effect of localizing particle 1, on
particle 2. In Strekalov's experiment, all the particles counted by D2
are such that the other particle of their pair has passed through the single
slit.  There are many pairs which are not counted, whose one member has
passed through the slit, but doesn't reach the fixed D1. However
as far as Popper's experiment is concerned, this is not important. As long as
the particles which are detected by D2 are those whose other partner passed
through the slit, they will show the effect that Popper was looking for.
Popper was inclined to predict that the test would decide
against the Copenhagen interpretation.

Let us look at the result of Strekalov et al's experiment (see Fig. 4).
The points represent the width of the diffraction pattern, in Strekalov et
al's experiment, as a function
of the slit width. For small slit width, the width of the diffraction
pattern sharply increases as the slit is narrowed. This is in clear
contradiction with Popper's prediction. To emphasize the point, we quote
Popper: \cite{popper}
\begin{quote}
``If the Copenhagen interpretation is correct, then such
counters on the far side of slit B that are indicative of a wide scatter
\ldots should now count coincidences; counters that did not count any
particles before the slit A was narrowed \ldots''
\end{quote}
Strekalov et al's experiment shows exactly that, if we replace the scanning
D2 by an array of fixed detectors.  So, we conclude that Popper's
test has decided in favor of Copenhagen interpretation.

\begin{figure}[h]
\centerline{\resizebox{4.5in}{!}{\includegraphics{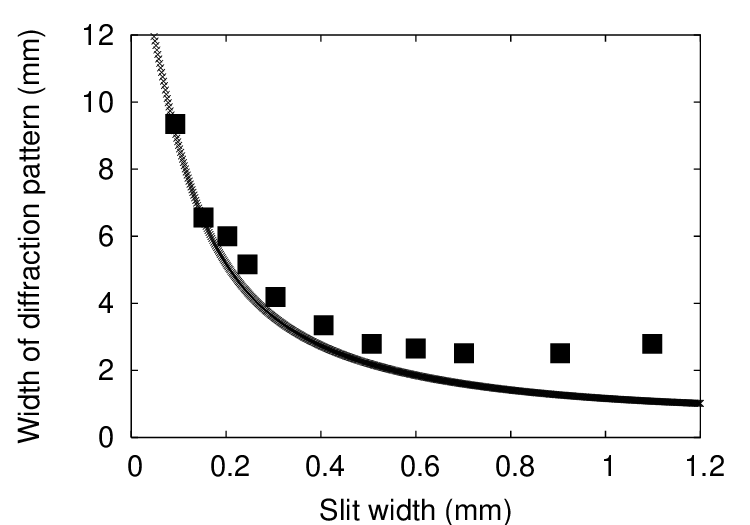}}}
\caption{Width of the diffraction pattern, plotted against the
full width of slit A.
The squares represent the data of Strekalov et al's experiment \cite{ghost}.
The line represents the theoretical width, calculated from
(\ref{width}) for $\hbar/\sigma=0.04$ mm,
using the parameters of Strekalov et al's experiment }
\end{figure}

The theoretical analysis carried out by us should apply to Strekalov et al's
experiment, with the understanding that the single slit interference
pattern is seen only if D1 is fixed.  In other words, if D1 were also
scanned along y-axis, the diffraction pattern would essentially remain the
same except that the smaller peaks, indicative of interference from
different regions within the slit, would be absent. We use (\ref{width})
to plot the full width at half maximum of the diffraction pattern
against, $2\epsilon$, which we assume to be the full width of the
rectangular slit A (see Fig 4). The plot uses $2L_1+L_2=1.8$ m, the value
used in Ref. \citen{ghost}, and an arbitrary $\hbar/\sigma=0.04$ mm. 
Our graph essentially agrees with that of Strekalov et al. Some deviation
is there because we have not taken into account the beam geometry, and the
finite size (0.5 mm) of the detectors, which will lead to an additional 
contribution to the width.

\section{Discussion and conclusion}

In 1987, when Collet and Loudon \cite{collet} argued that the use of a
stationary source was fundamentally flawed, the general view was that
Popper experiment will not be able to test the Copenhagen interpretation
of quantum mechanics. Short has also emphasized that the imperfect localization
is a manifestation of the problem pointed out by Collet and Loudon,
and concluded that the experiment cannot implement Popper's test \cite{short}.
Kim and Shih's experiment actually avoids this problem by obtaining a ghost
image of the slit.

We have shown that Strekalov et al's ghost interference experiment, actually
implements Popper's test in a conclusive way, but the result is in
contradiction with Popper's prediction.  It could not have been
otherwise, because our theoretical analysis shows that the results are
a consequence of the formalism of quantum mechanics, and not of any
particular interpretation. This was also pointed out by Krips, who
predicted that narrowing slit A would lead to increase in the width of
the diffraction pattern behind slit B (in coincident counting) \cite{krips}.
So, Krips prediction has been vindicated by Strekalov et al's experiment.

In our view, the only robust criticism of Popper's experiment was that
by Sudbery, who pointed out that in order to have perfect correlation 
between the two entangled particles, the momentum spread in the initial
state, had to be truly
infinite, which made any talk of additional spread, meaningless
\cite{sudbery,sudbery2}. For some reason, the implication of Sudbery's
point was not fully understood. It is this very point which, when
generalized, leads to our conclusion that no additional momentum spread
in particle 2 can be seen, even in principle.

Thus, our conclusion is that although Kim and Shih's experiment circumvents
the objections raised by Collet and Loudon, it is not conclusive about
Popper's test. On the other hand, Strekalov et al's experiment, implements
Popper's test in a conclusive way. Their results vindicate the Copenhagen
interpretation of quantum mechanics (if one takes Popper's viewpoint).
In reality, the results are just a manifestation of quantum mechanics, which
hardly needs any more vindication at this stage.

\section*{Acknowledgments}
The author acknowledges valuable suggestions from Virendra Singh and useful
discussions with Pankaj Sharan on narrowing wave-packets.

\end{document}